\documentclass[aps,prb,preprint,showpacs,groupedaddress,superscriptaddress]{revtex4-1}
\usepackage{amssymb,amsmath,bm}
\usepackage[dvips]{graphicx}
\usepackage{color}

\begin{document}

\title{Zero-field Hall effect in chiral $p$-wave superconductors near Kosterlitz-Thouless transition}
% Force line breaks with \\

\author{C. K. Chung}
\affiliation{Department of Physics, The University of Tokyo, 7-3-1 Hongo, Tokyo 113-0033, Japan}
\author{Y. Kato}
\affiliation{Department of Basic Science, The University of Tokyo, 3-8-1 Komaba, Tokyo 153-8902, Japan}
\date{\today}
\begin{abstract}
We discuss Hall effect and power dissipation in chiral $p$-wave superconductors near Kosterlitz-Thouless transition in the absence of applied magnetic field. In bound pair dynamics picture, nonzero Hall conductivity emerges when vortex-antivortex bound pair polarization has a component transverse to the direction of external perturbation. Such effect arises from the broken time reversal symmetry nature of a chiral $p$-wave superconducting state and does not require an applied magnetic field. A frequency-dependent matrix dielectric function $\epsilon(\omega)$ is derived to describe the screening effect due to the pair polarization. Quantities related to the Hall conductivity and power dissipation, denoted as $\epsilon^{-1}_\perp$ and $\Im (-\epsilon^{-1}_\parallel)$, are investigated in frequency and temperature domain. The imaginary part of the former can show peak structure and sign reversal as a function of frequency close to transition temperature, as well as in the temperature domain at various fixed frequencies. The latter shows peak structure near transition temperature. These features are attributed to pair-size-dependent longitudinal and transverse response function of bound pairs. Consequences due to free vortex dynamics and the resulting total conductivity tensor $\sigma$ are also discussed.
\end{abstract}

\maketitle

\section{Introduction}

Zero-field Hall effect in chiral $p$-wave superconductors (SCs) has drawn much attention in literature recently. \cite{Volovik:1, Furusaki:1, Goryo:1, Lutchyn:1, Taylor:1, Wysokinski:1} Because of the nature of broken time reversal ($\mathcal T$) symmetry, a nonzero Hall conductivity can be possible in a chiral $p$-wave SC. Indeed, it has already been shown that spontaneous Hall effect could arise from the intrinsic angular momentum of Cooper pairs \cite{Volovik:1} as well as from the spontaneous surface current. \cite{Furusaki:1} More recently, Hall conductivity due to impurity effect \cite{Goryo:1, Lutchyn:1} or to multiband SC structure \cite{Taylor:1, Wysokinski:1} was also studied, which could give possible explanation to the observed polar Kerr effect in the superconducting state of Sr$_2$RuO$_4$. \cite{Xia:1}

In this work, we address the zero-field Hall effect in a chiral $p$-wave SC originating from another mechanism, namely the vortex dynamics near Kosterlitz-Thouless (KT) transition. In two-dimensional (2D) superfluid (SF) or SC films, quantized vortices are realized as topological defects in the condensates, whose dynamics has been one of the key ingredients in understanding 2D phase transition phenomena. \cite{Barber:1} A few decades ago, Kosterlitz and Thouless \cite{Kosterlitz:1} suggested a static theory to relate a phase transition observed in superfluid $^4$He film \cite{Herb:1} to vortex-antivortex pair unbinding process across a transition temperature $T_\text{KT}$. In this picture, the logarithmic vortex-antivortex interaction is screened by smaller pairs and is renormalized to $1/\widetilde \epsilon$ of its bare value $K_0$ for temperature $T \leq T_\text{KT}$. The length-dependent dielectric constant $\widetilde \epsilon$ is used to describe the static screening of pair interaction. When $T>T_\text{KT}$, there exists a finite pair size $\xi_+$ such that the interaction becomes vanishingly small. Consequently, the pair unbinds and free vortices emerge; superfluidity is then destroyed. Soon after that, Ambegaokar, Halperin, Nelson, and Siggia (AHNS) \cite{Ambegaokar:1,Ambegaokar:2,Ambegaokar:3} combined this static theory with Hall and Vinen's dynamical description of vortex motion \cite{Vinen:1} to give an analysis of the dynamical effect on the phase transition. Concisely speaking, the renormalization process in the static theory \cite{Kosterlitz:2} is truncated by vortex dynamics with a characteristic length $r_D (\omega) = \sqrt{14D / \omega}$ instead of going to its completion. \cite{Ambegaokar:1} Here $D$ is the diffusivity of vortex movement and $\omega$ is the driving frequency. This results in broadening transition observed in $^4$He SF films \cite{Bishop:1} as well as in charged Fermi systems such as high-temperature SCs. \cite{Festin;Festin}

It is instructive to explore any physical consequence stemming from this KT transition in a broken $\mathcal T$ symmetry state. Possible experimental candidates of broken $\mathcal T$ symmetry state could be superconducting Sr$_2$RuO$_4$ (Ref.~\onlinecite{Maeno:1}) or $^3$He-A phase thin film, \cite{Volovik:2} in which pairing of chiral $p$-wave type is expected. Indeed, in literature some theoretical works have been done to investigate new features specific to SCs with pairing of this type near KT transition. \cite{Herland;Bauer}

In this work, we consider a 2D $p$-wave pairing state with $d$-vector $ \boldsymbol d = \hat z (p_x \pm i p_y)/p_\text{F}$ where $\hat z$ is the unit vector normal to film surface, $p_x $ and $p_y$ denote the $x$ and $y$  component of the relative momentum $\boldsymbol p$ of a Cooper pair, and $p_\text{F}$ is the Fermi momentum. Assuming isotropic Fermi surface, two kinds of pairing fields can be obtained:
$
\Delta^{(+-)} (\bm \rho,\bm p)/\Delta_\infty
=
f_1^{(+-)}(\rho) e^{-i\phi} (p_x + i p_y)/p_\text{F}
+ f_2^{(+-)}(\rho) e^{i\phi} (p_x - i p_y)/p_\text{F}
$
 and
$
\Delta^{(++)} (\boldsymbol r,\boldsymbol p)/\Delta_\infty
=
f_1^{(++)}(\rho) e^{i\phi} (p_x + i p_y)/p_\text{F}
+ f_2^{(++)}(\rho) e^{3i\phi} (p_x - i p_y)/p_\text{F}
$
with asymptotic behavior at large $\rho$ being
$
f_1^{(+-)} \rightarrow 1
$, $
f_2^{(+-)} \rightarrow 0
$, $
f_1^{(++)} \rightarrow 1
$, and $
f_2^{(++)} \rightarrow 0
$. \cite{Heeb:1, Matsumoto:1, Kato:2}  Here $\bm \rho = \rho (\cos \phi , \sin \phi)$ is the spatial coordinate and $\Delta_\infty$ is the modulus of the energy gap in the bulk. From these pairing fields, we can identify two types of integer vortices called $(+-)$--vortex and $(++)$--vortex respectively.

Because of spontaneously broken $\mathcal T$ symmetry, these two types of vortices are not equivalent. \cite{Kato:1, Kato:2, Kato:3} In particular, their Hall and Vinen coefficients \cite{Vinen:1} do not share the same value, i.e., $B^{(++)} \neq B^{(+-)}$ and ${B^\prime}^{(++)} \neq {B^\prime}^{(+-)}$ (see section II A). This results in a nonzero ``convective" term in a vortex pair polarization Fokker-Planck equation in addition to the conventional diffusive terms, while in its $s$-wave counterpart such convective motion does not enter the dynamics. \cite{Ambegaokar:2} The relative strength of convection is quantified by a convective ratio $x_0$ in this paper. It is due to such distinct feature that pair polarization transverse to the driving force field becomes possible even without applied magnetic field. A nonzero vortex-dynamics-induced Hall conductivity $\sigma_\perp$ then follows naturally.

%The Hall conductivity is found to be nonzero given finite a $x_0$.

The main result of this work is that in the bound pair dynamics description, we obtain nonvanishing Hall conductivity $\sigma_\perp$ and ac conductivity $\sigma_\parallel$ near the KT transition. One of the interesting features in the Hall conductivity is that strong positive peak and sign changes in $\omega \Re(\sigma_\perp)$ are observed at suitable frequency region above the transition temperature, as well as above $T_\text{KT}$ in temperature domain at fixed frequencies. On the other hand, $\omega \Re(\sigma_\parallel)$ is shown to have similar features as in AHNS's results. We note that the shapes of two length-dependent response functions $\mathcal G_{\parallel}(r,\omega)$ and $\mathcal G_{\perp}(r,\omega)$, which corresponds respectively to the longitudinal and transverse response of bound pairs with separation $r$ to external perturbation with frequency $\omega$, play a determining role on the behavior of $\sigma_\perp$ and $\sigma_\parallel$. We also discuss the contribution of free vortex motion and the resulting total conductivity tensor.

The paper is organized as follows: In section II, we generalize AHNS's vortex dynamics in the chiral $p$-wave context. To describe the vortex-antivortex bound pair dynamics, the above-mentioned response functions $\mathcal G_{\parallel}(r,\omega)$ and $\mathcal G_{\perp}(r,\omega)$ are derived from the Fokker-Planck equation governing the pair motion. Together with the free vortex contribution, we arrive at a matrix dielectric function $\epsilon (\omega)$ which describes the total screening effect under time-dependent perturbation. In section III, we investigate the frequency and temperature dependence of the conductivity tensor $\sigma(\omega)$ constructed from $\epsilon (\omega)$, treating the bound pair and the free vortex contribution separately. The behavior of total conductivity $\sigma$ is also discussed. A summary and remark are given in section IV. Finally, we discuss analytic expression of $\Im (-\epsilon^{-1}_\parallel)$, $\Im(\epsilon^{-1}_\perp)$, and $\Re(-\epsilon^{-1}_\perp)$ in opposite limit of the convective ratio $x_0\ll1$ and $x_0\gg 1$ in Appendix A.

\section{Matrix dielectric function}

\subsection{Equations of motion for single vortex}
To construct a matrix dielectric function, we consider a neutral SF film system resembling that employed in AHNS's dynamical theory, with film thickness of order the superconducting coherence length and linear dimension $L$ ($W$) along $x$ ($y$) direction. $L$ is very large and $W$ is large but finite. Vortex core motion relative to the local superfluid velocity leads to a Magnus force $\bm F_\text{M}$ \cite{Ambegaokar:3}
\begin{eqnarray}
\boldsymbol {F}_\text{M}^{(i)} &=&
n_i \rho_\text{s}^0 \frac{2\pi \hbar}{m^\ast} \hat z \times
\left(\boldsymbol v_\text{L}^{(i)} - \boldsymbol{v}_\text{s}^{(i)} \right).
\end{eqnarray}
In the above equation, $\boldsymbol v_\text{L}^{(i)}$ and $\boldsymbol{v}_\text{s}^{(i)}$ are the velocity of the $i$-th vortex core and the local superfluid flow excluding the diverging self-field of the $i$-th vortex respectively. $n_i (=\pm 1)$ is the vorticity of the $i$-th vortex. $\rho_\text{s}^0$ is the bare areal superfluid mass density, which is defined as the three-dimensional superfluid density integrated across the film thickness. $m^\ast$ is the mass of the constituting particle, which is equal to the mass of a Cooper pair.

The SF film is driven by a vibrating substrate. A vortex core moving relatively to the substrate experiences a vorticity-dependent drag force $\boldsymbol{F}_\text{D}^{(i)}$ \cite{Ambegaokar:3}
\begin{eqnarray}
\boldsymbol{F}_\text{D}^{(i)} &=&
{B}^{(i)} \left( \boldsymbol{v}_\text{n} - \boldsymbol v_\text{L}^{(i)} \right)
+
{B^\prime}^{(i)} n_i \hat z \times \left( \boldsymbol{v}_\text{n} - \boldsymbol v_\text{L}^{(i)} \right), \label{FD}
\end{eqnarray}
where $\boldsymbol{v}_\text{n}$ is the moving substrate velocity. ${B}^{(i)}$ and ${B^\prime}^{(i)} $ denote the vorticity-dependent drag coefficients originating from interactions with the substrate and with thermally excited quasiparticles and collective modes.
Here $B^{(i)}$ (${B^\prime}^{(i)}$) $= B^{(++)}$ (${B^\prime}^{(++)}$ ) for $n_i = 1$ or $B^{(+-)}$ (${B^\prime}^{(+-)}$) for $n_i = -1$. These quantities have been obtained for a three-dimensional clean $s$-wave SF/SC with isotropic Fermi surface, \cite{Kopnin:1} and are related to the relaxation time $\tau$ for the Caroli-deGennes-Matricon mode \cite{Caroli:1} in the vortex core. Due to broken $\mathcal T$ symmetry, the relaxation time $\tau^{(i)}$ for the mode in the $(++)$--vortex and $(+-)$--vortex core, and thus the values of their drag coefficients, are different, resulting in the vorticity-dependent drag force $\boldsymbol{F}_\text{D}^{(i)}$ on the vortex core. These drag coefficients for 2D clean SF/SCs with cylindrical Fermi surface can be inferred from the three-dimensional result easily.

When two forces balance $\boldsymbol{F}_\text{D}^{(i)} + \boldsymbol{F}_\text{M}^{(i)}=0$, the $i$-th vortex velocity is expressed by
\begin{eqnarray}
\bm v_\text{L}^{(i)} &=&
n_i 2\pi K_0 D^{(i)}\frac{ m^\ast}{\hbar}  \hat z \times (\bm {v}_\text{n} - \bm {v}_\text{s}^{(i)})
+
C^{(i)} (\bm {v}_\text{n} - \bm {v}_\text{s}^{(i)})
+
\bm {v}_\text{s}^{(i)} + \bm {\eta}^{(i)} (t), \label{one_vortex}
\end{eqnarray}
where
\begin{eqnarray}
D^{(i)}
&=&
k_\text{B}T \frac{B^{(i)}}{(2\hbar \pi \rho_\text{s}^0  / m^\ast - {B^\prime}^{(i)})^2 + {B^{(i)}}^2},
\\
C^{(i)}
&=&
1- \frac{2\hbar \pi \rho_\text{s}^0 }{m^\ast}
\frac{2\hbar \pi \rho_\text{s}^0  / m^\ast - {B^\prime}^{(i)}}
{(2\hbar \pi \rho_\text{s}^0  / m^\ast - {B^\prime}^{(i)})^2 + {B^{(i)}}^2 }.
\end{eqnarray}
Here $K_0 = \hbar^2 \rho_s^0/({m^\ast}^2 k_\text{B} T)$ with $k_\text{B}$ being the Boltzmann constant. $\bm {\eta}^{(i)} (t)$ are fluctuating Gaussian noise sources incorporated to bring the vortices to equilibrium. Their components satisfy
$
\langle \eta^{(i)}_\mu (t) \eta^{(j)}_\nu (t^\prime) \rangle
=
2D^{(i)} \delta_{ij} \delta_{\mu \nu} \delta (t-t^\prime)
$.
From Ref.~\onlinecite{Ambegaokar:3}, the local superfluid velocity $\bm {v}_\text{s}^{(i)}$ is related to the spatial average superfluid velocity $\bm  u_\text{s}$ and the positions of individual vortices $\bm  r_j$ by
$
\bm {v}_\text{s}^{(i)}
=
\bm {u}_\text{s}
+
(\hbar/ m^\ast) \hat z \times
\sum_{j \neq i} n_j \nabla \mathfrak G(\bm {r}_i,\bm {r}_j)
$. Here $\mathfrak G(\bm {r},\bm {r}_j)$ is the Green function satisfying
$
\nabla^2 \mathfrak G (\bm {r},\bm {r}_j)
= 2\pi \Delta(\bm {r}-\bm {r}_j)
$
and boundary condition $\mathfrak G(\bm {r},\bm {r}_j)=0$ on the edges. The function $\Delta(\bm {r}-\bm {r}_j) $ is localized in a region around $\bm {r}_j$ with radius of the order of coherence length (we could say that the function $\Delta(\bm {r}-\bm {r}_j) $ is a delta function in the coarse-grained scale). Far away from the edges, $\mathfrak G(\bm {r},\bm {r}_j) \approx \ln (\vert \bm {r} - \bm {r}_j \vert/a_0)$. The spatial average of $\nabla \mathfrak G$ turns out to be zero and thus $\bm {u}_\text{s} $ represents the spatial average of $\bm {v}_\text{s}^{(i)}$. Finally, the time evolution of $\bm {u}_\text{s} $ obeys
$
\partial_t \bm {u}_\text{s}
=
- \sum_i n_i (2\pi \hbar) / (LW m^\ast) \hat z \times
\bm v_\text{L}^{(i)},
$
reflecting the fact that the average superfluid velocity in the $x$ direction changes by $2\pi \hbar/(L m^\ast)$ when a vortex with $n_i=1$ moves across a strip with width $\delta y$.

In a charged system, if we follow Kopnin's description, \cite{Kopnin:3} the driving force on the vortex due to a transport current
$\bm  j_\text{tr}$ is a Lorentz force
$\bm  F_\text{L}^{(i)}  = n_i \Phi_0[\bm  j_\text{tr} \times \hat z \text{ sgn} (e)]/c $ where
$\Phi_0 = \hbar \pi c/\vert e \vert$ and $e$ is the electron charge. This force is balanced by the force from environment
$\bm  F_\text{env}^{(i)} = -\eta^{(i)}  \bm  v_\text{L}^{(i)} +n_i {\eta^\prime}^{(i)} \hat z \times \bm  v_\text{L}^{(i)}$. If we set $\bm  v_\text{n}=0$ in Eqs.~(\ref{FD}) and (\ref{one_vortex}), the results derived in a neutral system can be carried over to a charged system by the translation
$\eta^{(i)}\leftrightarrow B^{(i)} $ and
$ {\eta^\prime}^{(i)}\leftrightarrow 2\pi \hbar \rho_s^0/m^\ast - {B^\prime}^{(i)}$.

\subsection{Fokker-Planck equation for bound pair contribution}

We consider the polarization of a test vortex-antivortex pair whose constituting vortices interact via a screened logarithmic interaction. The pair is under the influence of an infinitesimal oscillating external field $\delta \bm  {E}$. The Langevin equation for their relative coordinate $\bm  {r}$ can be obtained by subtracting Eq.~(\ref{one_vortex}) from each other for opposite vorticity
\begin{eqnarray}
\frac {d \bm  {r}}{dt}
&=&
-  4 \pi \overline D K_0 \nabla U(\bm  {r})  +2\overline C \hat z \times \nabla U(\bm  {r}) + \bm  {\eta}. \label{pair_EOM}
\end{eqnarray}
Here $\overline D =  (D^{(++)} + D^{(+-)})/2$ and $\overline C =  (C^{(++)} - C^{(+-)})\hbar/(2m^\ast)$. $U(\bm {r})$ is one-half the dimensionless potential energy of the pair and the Gaussian noise now satisfies
$
\langle \eta_\mu \eta_\nu \rangle
=
4 \overline D\delta_{\mu \nu} \delta (t-t^\prime)
$. The potential energy is given by
\begin{eqnarray}
U(\boldsymbol {r})&=& \int_{a_0}^r dr \frac{1}{r \widetilde \epsilon (r)} - \frac{\mu_0 {m^\ast}^2}{\pi \hbar^2 \rho_s^0} - \frac{m^\ast}{\hbar} \delta \boldsymbol {E} \cdot \boldsymbol r.
\end{eqnarray}
In the above equation, the first term on the right hand side describes the logarithmic interaction screened by the Kosterlitz dielectric constant $\widetilde \epsilon(r)$. $2\mu_0$ in the second term is related to the energy required to create a pair with separation $a_0$. In the last term, $\delta \boldsymbol E$ has dimension of velocity and acts as the perturbation. In the integration limit $a_0$ is a length scale related to the size of a vortex core, and $r$ is the pair separation.

We can see in Eq.~(\ref{pair_EOM}) that in addition to the conventional diffusive terms depending on $\overline D$, a convective term proportional to $\overline C$ also enters the dynamical equation. While the strength of the former is proportional to the \emph{average} of $D^{(i)}$, that of the latter is related to the \emph{difference} of $C^{(i)}$ between opposite vorticity. We emphasize here that such a convective pair motion is one of the special features for a system with unequal opposite vortices, and is thus absent in an $s$-wave SF/SC since $C^{(+)} = C^{(-)}$ in that case. Given this nonzero $\overline C$, the pair polarization is tilted away from the direction of the force field $-\nabla U(\boldsymbol {r})$, and has both longitudinal and transverse components even without applied magnetic field.

The Fokker-Planck equation corresponding to Eq.~(\ref{pair_EOM}) is given by
\begin{eqnarray}
\frac {\partial \Gamma(\boldsymbol r,t)}{\partial t} &=& 4 \pi \overline D K_0 \nabla \cdot (\Gamma \nabla U ) -
2\overline C \nabla \cdot (\Gamma \hat z \times \nabla U )
+ 2 \overline D \nabla^2 \Gamma,
\end{eqnarray}
where $\Gamma(\boldsymbol r,t)$ is the density of pairs per unit area of separation. We take the time-independent state $\Gamma_0$ to be $\exp ({-2\pi K_0 U_0})/a_0^4$ where $U_0 = U\vert_{\delta \boldsymbol {E}=0}$. Now we follow the standard procedure, \cite{Ambegaokar:2} letting $\Gamma= \Gamma_0 + \delta \Gamma$ and keeping terms to first order in $\delta \boldsymbol E $. In frequency space, we obtain
\begin{eqnarray}
-i\omega \delta \Gamma (\boldsymbol r,\omega)
&=&
-4 \pi \overline D K_0 \nabla \cdot
\left[
\left( \frac{m^\ast}{\hbar} \delta \boldsymbol E \Gamma_0 \right)
-
\delta \Gamma \nabla U_0
\right]
\nonumber\\ &&
+
2 \overline C \nabla \cdot
\left[
\left( \frac{m^\ast}{\hbar} \hat z \times \delta \boldsymbol E \Gamma_0 \right)
-
\delta \Gamma \hat z \times \nabla U_0
\right]
+
2 \overline D \nabla^2 \delta \Gamma.
\end{eqnarray}
We employ the expansion $\delta \Gamma (\boldsymbol {r},\omega)= \sum_m \delta \Gamma_m (r,\omega) e^{i m\theta}$, where $\theta$ is the angle measured from $\delta \boldsymbol {E} / \vert \delta \boldsymbol {E} \vert$ to $\hat r$ in anti-clockwise sense. Only $\delta \Gamma_m$ with $m=1$ and $m=-1$ are coupled to the external field.

We define a convective ratio $x_0 = \overline C/(2\pi K_0 \overline D)$, which measures the relative strength between convection and diffusion, and introduce an ansatz
$
\delta \Gamma_{\pm 1} (r,\omega)  = 2 \pi  K_0 \left( 1 \pm i x_0  \right) \delta \boldsymbol {E} \Gamma_0$ $ g_\pm (r,\omega) r m^\ast/ (2\hbar)$. Together with an approximation \cite{Ambegaokar:2} $2\pi K_0/ \widetilde \epsilon =4 $ (which is valid near the transition) and a change of variable $z^2 = -i\omega r^2 /(2 \overline D)$, the equations of $g_\pm(z)$ in the ansatz corresponding to angular momentum $m=\pm1$ are given by
\begin{eqnarray}
0&=& -\left(z^2+ 4 \pm 4i x_0\right) g_\pm(z)- zg_\pm^\prime (z) + z^2 g_\pm^{\prime\prime}(z) +4. \label{eqt_for_g}
\end{eqnarray}
We are then able to write down the change of distribution function $\delta \Gamma = \delta \Gamma_1 e^{i\theta} + \delta \Gamma_{-1} e^{-i\theta}$ in terms of $ \delta \boldsymbol E $ explicitly
\begin{eqnarray}
\delta \Gamma (\boldsymbol {r},\omega) &=&  2\pi K_0 \Gamma_0 \frac{m^\ast}{\hbar}
\left[
\delta \boldsymbol E \cdot \boldsymbol r
\mathcal{G}_\parallel(r,\omega)
-
\delta \boldsymbol E \cdot (\boldsymbol r \times \hat z)
\mathcal{G}_\perp(r,\omega)
\right], \label{d_Gamma}
\end{eqnarray}
where
\begin{eqnarray}
\mathcal{G}_\parallel(r,\omega)
=
\frac{1}{2}\left[ G_+ (r,\omega) + G_- (r,\omega) \right],
\quad\quad
\mathcal{G}_\perp(r)
=
-\frac{i}{2}\left[ G_+ (r,\omega) - G_- (r,\omega) \right], \label{G_g}
\end{eqnarray}
and $ G_\pm (r,\omega) = ( 1 \pm ix_0) g_\pm (r,\omega)$. In Eq.~(\ref{d_Gamma}), $\mathcal{G}_\parallel(r,\omega)$ takes the role of pair-size-dependent response longitudinal to the driving field $\delta \boldsymbol E $ and $\mathcal{G}_\perp(r,\omega)$ is the transverse response function. In the limit of vanishing $x_0$, $\mathcal{G}_\parallel(r,\omega)$ reduces to the $s$-wave SF/SC result $g_s(r, \omega) \approx 14D/(14D - i \omega r^2)$ while $\mathcal{G}_\perp(r,\omega)$ becomes identically zero. In Eq.~(\ref{G_g}), we find that $\mathcal{G}_{\parallel,\perp}(r,\omega)$ depend on two quantities $G_\pm (r,\omega)$ which describe pair motion with angular momentum $m = \pm 1$ respectively.

\begin{figure}[htbp]
\includegraphics [width=0.8\textwidth] {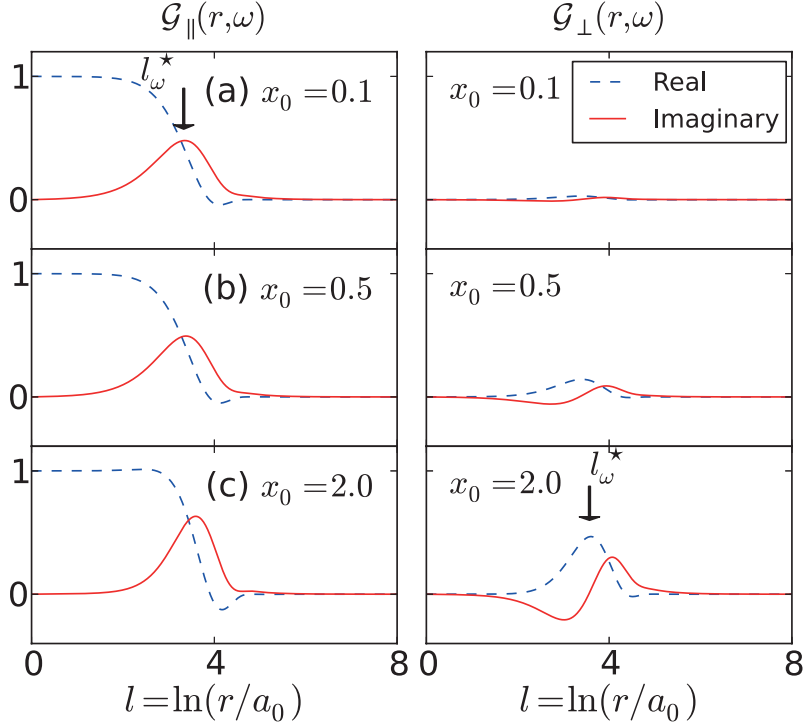}
\caption{(Color online) The longitudinal and transverse response function $\mathcal{G}_\parallel(r,\omega)$ (left column) and $\mathcal{G}_\perp(r,\omega)$ (right column) are plotted as a function of $l=\ln(r/a_0)$ for $x_0 =$ (a) $0.1$, (b) $0.5$, and (c) $2.0$ respectively. The real and imaginary part are indicated by blue dashed lines and red solid lines respectively. $\mathcal{G}_\parallel(r,\omega)$ is qualitatively similar to the $s$-wave SF/SC result. $\mathcal{G}_\perp(r,\omega)$ is a new feature with a peak in the real part and a dip-recouping shape in the imaginary part around $l_\omega^\star$.}
\label{mathcalG}
\end{figure}

We describe $\mathcal{G}_{\parallel,\perp}(r,\omega)$ first. In Fig.~\ref{mathcalG} we plot $\mathcal{G}_\parallel(r,\omega)$ and $\mathcal{G}_\perp(r,\omega)$ as a function of $l=\ln(r/a_0)$ for fixed frequency $\omega a_0^2/(2\overline D) =10^{-2}$ for (a) $x_0=0.1$, (b) $x_0=0.5$, (c) $x_0=2$. Blue dashed lines and red solid lines represent their real and imaginary parts respectively. In the left column of Fig.~\ref{mathcalG}, the longitudinal response function $\mathcal{G}_\parallel(r,\omega)$ is qualitatively similar to the $s$-wave SF/SC result $g_s(r, \omega)$ even for finite $x_0$. It has a step-function-like real part and a delta-function-like imaginary part concentrating near $l_\omega^\star$ ($l_\omega^\star$ marks the position where $\Im (\mathcal{G}_\perp)$ changes sign; see Appendix A). The response function $\mathcal{G}_{\perp} (r,\omega)$ is a new feature in this model. In the right column of Fig.~\ref{mathcalG}, $\Re [\mathcal{G}_\perp(r,\omega)]$ has a peak structure. This means that neither smaller nor larger pair gives response in transverse direction. Only pairs with characteristic pair size $l_\omega^\ast$ can give rise to the Hall effect. Besides, $\Im [\mathcal{G}_\perp(r,\omega)]$ shows negative dip shape for $l<l_\omega^\star$, and then recoups when $l>l_\omega^\star$. It turns out in later section that such a dip-recouping antisymmetric shape in $\Im (\mathcal{G}_\perp)$ about $l_\omega^\star$ plays a determining role in many features of the Hall conductivity. These features become more significant when $x_0$ increases.

\begin{figure}[htbp]
\includegraphics [width=0.8\textwidth] {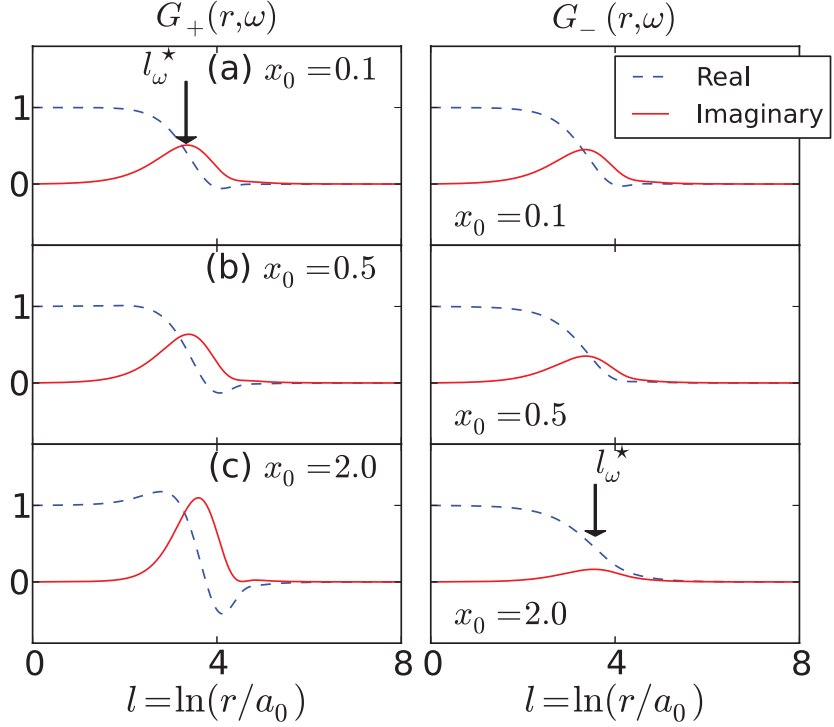}
\caption{(Color online) The response function $G_+(r,\omega)$ (left column) and $G_-(r,\omega)$ (right column) corresponding to angular momentum $m=\pm1$ motion are plotted as a function of $l=\ln(r/a_0)$ for $x_0 =$ (a) $0.1$, (b) $0.5$, and (c) $2.0$ respectively. The real and imaginary part are indicated by blue dashed lines and red solid lines respectively. Asymmetry between the two columns becomes more prominent when $x_0$ increases.}
\label{capitalG}
\end{figure}

In Eq.~(\ref{G_g}), we can see that $\mathcal{G}_{\parallel}(r,\omega)$ depends on the average of $G_\pm (r,\omega)$, i.e., the average response with $+1$ and $-1$ angular momentum, and $\mathcal{G}_{\perp}(r,\omega)$ is related to the \emph{difference} between them. This means that any asymmetry between $m=\pm1$ motion gives rise to nonzero transverse response. In Fig.~\ref{capitalG} we plot $G_\pm (r,\omega)$ as a function of $l$ using the same set of $\omega$ and $x_0$ as in Fig.~\ref{mathcalG}. Again, blue dashed lines and red solid lines represent their real and imaginary parts respectively. We observe that asymmetry between $m=\pm1$ motion grows with increasing $x_0$. For small $x_0$ in Fig.~\ref{capitalG}(a), the motion described by $G_+ (r,\omega)$ and $G_- (r,\omega)$ is only slightly asymmetric, and they look like the familiar curve $g_s( r, \omega)$. When $x_0$ increases in Fig.~\ref{capitalG}(b) and (c), the asymmetry between the left and right column becomes more and more significant. It is worth noting that, when we subtract from each other the real part of the functions for opposite $m$, we can obtain the dip-recouping shape of $\Im(\mathcal{G}_{\perp})$.

Equation (\ref{eqt_for_g}) can be solved exactly to give the length-dependent response functions, but here we can employ approximate solution to $g_\pm(r,\omega)$ in a manner similar to Ref.~\onlinecite{Ambegaokar:2} by neglecting $g_\pm^{\prime\prime}$ and $g_\pm^{\prime}$ in Eq.~(\ref{eqt_for_g}). Then $\mathcal{G}_{\parallel,\perp}$ become
\begin{eqnarray}
\mathcal{G}_\parallel(r,\omega)
&\approx&
\frac{1}{2} \left( \frac{1+ix_0}{1 - i  \omega r^2/\lambda \overline D + i x_0}
+
\frac{ 1-ix_0}{1 - i  \omega r^2/\lambda \overline D - i x_0} \right), \label{app_G1}
\\
\mathcal{G}_\perp(r,\omega)
&\approx&
-\frac{i}{2} \left( \frac{1+ix_0}{1 - i  \omega r^2/\lambda \overline D + i x_0}
-
\frac{ 1-ix_0}{1 - i  \omega r^2/\lambda \overline D - i x_0} \right), \label{app_G2}
\end{eqnarray}
where $\lambda$ is a factor to fit the exact curve. It is selected to be $14$ when $x_0=0$ and is of order $10$ for a wide range of $x_0$. From Eqs.~(\ref{app_G1}) and (\ref{app_G2}) we can see that for a small convective ratio $1>x_0>0$, there are poles given by $1 - i \omega r^2 /(\lambda \overline D) \pm i x_0 = 0$. This means that the pair size with which a pair gives a strong response differs from the standard result $r_D= \sqrt{14D/ \omega}$ by a length of order $\pm x_0 \sqrt{\lambda \overline D / \omega}$ for motion with $m=\pm1$, creating the asymmetry demonstrated in Fig.~\ref{capitalG}. For a large convective ratio $x_0>1$, such pole is removed for $m=-1$. Therefore, a broad and flat curve is expected [Fig.~\ref{capitalG}(c) right column].

\subsection{Matrix susceptibility $\chi_\text{b}$ for bound pair contribution}

In this subsection, we introduce a susceptibility matrix $\chi_\text{b}$ for the bound pair polarization
\begin{eqnarray}
\chi_\text{b} &=&
\begin{pmatrix}
\chi_\text{b}^{\parallel} & \chi_\text{b}^{\perp}
\\
-\chi_\text{b}^{\perp}& \chi_\text{b}^{\parallel}
\end{pmatrix},
\end{eqnarray}
whose components are defined as
\begin{eqnarray}
\chi_\text{b}^{\parallel}
=
\frac{2 \pi \hbar }{m^\ast } \int d^2 r
\left(
\frac{\boldsymbol r}{2} \cdot  \frac{\delta \Gamma}{\delta \boldsymbol E  }
\right),
\quad\quad
\chi_\text{b}^{\perp} =
\frac{2 \pi \hbar }{m^\ast } \int d^2 r
\hat z \cdot
\left(\frac{\boldsymbol r}{2}
\times
\frac{\delta \Gamma}{\delta \boldsymbol E }
\right). \label{chib}
\end{eqnarray}
They are so defined that the real parts of $\chi_\text{b}^{\parallel}$ and $\chi_\text{b}^{\perp}$ are positive when $x_0>0$. Using Eqs.~(\ref{d_Gamma}) and (\ref{chib}), we arrive at the expression
\begin{eqnarray}
\chi_\text{b}^{\parallel}
=
\int_{a_0}^{\xi_+} dr \frac{d \widetilde \epsilon}{dr}
\mathcal{G}_\parallel(r),
\quad\quad
\chi_\text{b}^{\perp}
=
\int_{a_0}^{\xi_+} dr \frac{d \widetilde \epsilon}{dr}
\mathcal{G}_\perp(r). \label{chibb}
\end{eqnarray}
The infinitesimal external field redistributes the pair polarization by an amount $\delta \bm P_\text{b}$ according to $ \delta \bm P_\text{b} =\chi_\text{b} \delta \boldsymbol E $. The definition in Eq.~(\ref{chib}) means the external field tilts the polarization in clockwise direction when $x_0>0$. It should also be noted that the integration are performed from ${a_0}$ to $\xi_+$ where $\xi_+$ is a coherence length with behavior $\xi_+ \rightarrow \infty$ for $T \leq T_\text{KT}$ and $\xi_+ \sim \exp [2\pi /(b \sqrt{T/T_\text{KT}-1})]$ for $T>T_\text{KT}$. $b$ is a non-universal positive constant of order unity. From Eq.~(\ref{chibb}) it is now clear that while the longitudinal polarization $\mathcal{G}_\parallel$ takes the role of the function $g_s(r,\omega)$ in AHNS's theory, the transverse polarization $\mathcal{G}_\perp$ has no simple analogy in $s$-wave SF/SCs and is specific to systems with finite convective ratios. Integrating the two pair-size-dependent response functions with the weight function $d \widetilde \epsilon / dr$ gives the susceptibility matrix elements from the bound pair contribution.

\subsection{Free vortex contribution and total dielectric function}

As for the free vortex contribution, from Eq.~(\ref{one_vortex}) we can obtain the equation for polarization $\bm P_\text{f}$ for plasma of free vortices under the spatial averaged driving field $\langle \boldsymbol E \rangle$. The total free vortex density is given by $n_\text{f} = N/(LW)$ with $N$ being the total number of free vortices and $LW$ being the area of the film. In frequency space, we have
\begin{eqnarray}
-i \omega \bm P_\text{f} &=&     \frac{2\pi \hbar}{m^\ast} n_\text{f}
\left[
\frac{2\pi \overline  D \hbar}{m^\ast}\frac{\rho_s^0}{k_B T}  \langle \boldsymbol E \rangle
-
\frac{m^\ast}{\hbar} \overline C \hat z \times \langle \boldsymbol E \rangle
 \right].
\end{eqnarray}
If we define the susceptibility for free vortices as $\bm P_\text{f} = \chi_\text{f} \langle \boldsymbol E \rangle$, we have
\begin{eqnarray}
\chi_\text{f} &=&
\frac{\gamma_0}{-i\omega}
\begin{pmatrix}
1 & x_0
\\
-x_0 & 1
\end{pmatrix}, \label{free}
\end{eqnarray}
where $\gamma_0 =  4 \pi^2 n_\text{f} \overline D K_0$. On the other hand, the free vortex density is related to the coherence length by $n_\text{f} = 2\pi C_1 \xi_+^{-2}$ where $C_1$ is a positive constant of order unity. \cite{Halperin:1}

Together with the bound pair contribution, the total dielectric function can be obtained as
$
\epsilon (\omega)= 1  + \chi_\text{b}+ \chi_\text{f}
$
and its inverse reads
\begin{eqnarray}
\epsilon^{-1} (\omega)
&=&
\begin{pmatrix}
\epsilon^{-1}_\parallel & \epsilon^{-1}_\perp
\\
-\epsilon^{-1}_\perp & \epsilon^{-1}_\parallel
\end{pmatrix},
\end{eqnarray}
where
$\epsilon^{-1}_\perp = -(\chi_\text{b}^\perp  + \chi_\text{f}^\perp)/ [(1  + \chi_\text{b}^\parallel +\chi_\text{f}^\parallel)^2 + (\chi_\text{b}^\perp  + \chi_\text{f}^\perp )^2]$ and
$\epsilon^{-1}_\parallel = (1  + \chi_\text{b}^\parallel + \chi_\text{f}^\parallel)/[(1  + \chi_\text{b}^\parallel +\chi_\text{f}^\parallel)^2 + (\chi_\text{b}^\perp  + \chi_\text{f}^\perp )^2]$. We are now in position to present our results on the Hall conductivity and power dissipation due to this dielectric function.

\section{Result and discussion in charged system}

It was discussed that the conductivity tensor $\sigma$ in a charged system was related to the inverse dielectric function $\epsilon^{-1}$ discussed above. \cite{Halperin:1} We can understand the relation by considering the total current under the influence of a driver coil electric field $\mathcal {E}$. \cite{Minnhagen:1} The vortex-modified total current $\bm j_\text{tot}$ is related to the field by $\mathcal {E} = -i \omega L_k \epsilon \bm j_\text{tot}$ where $L_k$ is the sheet kinetic inductance. This relation is generalized to our model with a matrix dielectric function $\epsilon$. If we use the sign convention that the normal state electron Hall conductivity with magnetic field pointing in $\hat z$ direction is positive, $\sigma(\omega)$ and $\epsilon^{-1}(\omega)$ is related by
\begin{eqnarray}
\sigma (\omega) &=&
\begin{pmatrix}
\sigma_\parallel & -\sigma_\perp
\\
\sigma_\perp & \sigma_\parallel
\end{pmatrix}
= \frac{ 1}{-i \omega L_k} \epsilon^{-1} (\omega).
\end{eqnarray}
In particular, we may write down
$
\omega L_k \Re ( \sigma_\perp) =  \Im (\epsilon_\perp^{-1})
$, $
\omega L_k \Im ( \sigma_\perp) =  \Re (-\epsilon_\perp^{-1})
$, and $
\omega L_k \Re ( \sigma_\parallel) =  \Im (-\epsilon_\parallel^{-1})
$. Among these expressions, the first two are associated with the real and imaginary part of the Hall conductivity, while the last one is related to the power dissipation $P$ because $P \propto \Re(\sigma_\parallel)\propto \Im (-\epsilon^{-1}_\parallel)/\omega$.

In the following subsections, we investigate the frequency and temperature dependence of $\Im (\epsilon_\perp^{-1})$, $\Re (-\epsilon_\perp^{-1})$, and $\Im (-\epsilon_\perp^{-1})$ respectively.

\subsection{Hall conductivity from bound pair contribution}

\begin{figure}[htbp]
\includegraphics [width=0.8\textwidth] {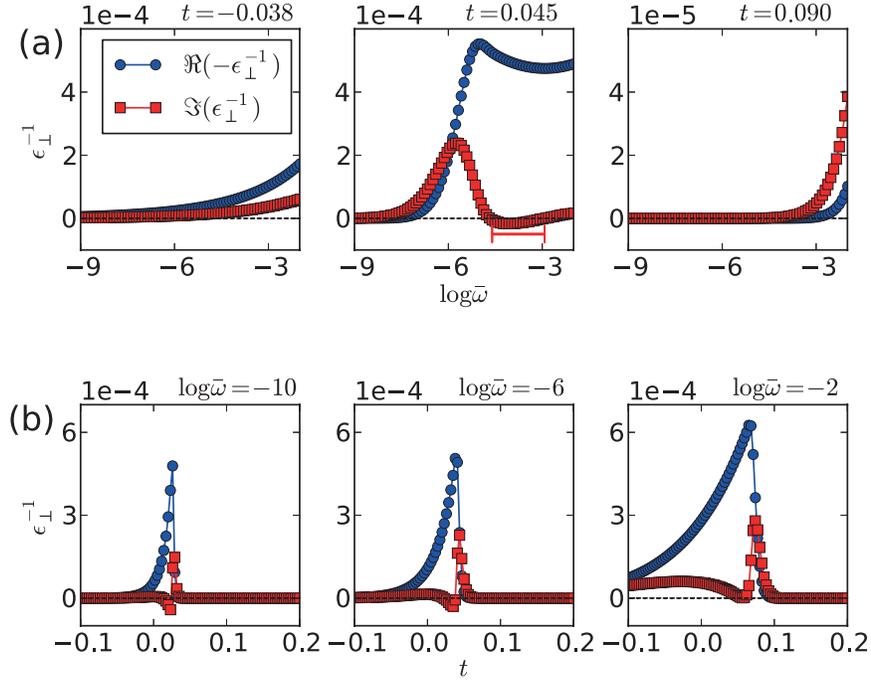}
\caption{(Color online) Bound pair contribution. The negative real part (blue circles) and imaginary part (red squares) of $\epsilon_\perp^{-1}$ as a function of (a) $\log \overline \omega$ and (b) $t=T/T_\text{KT}-1$. The other parameters used are $x_0=0.1$, $2\pi Y_0=0.1$ and $X_0^{(\text{KT})} \approx-0.215$. In (a), above the transition temperature, $\Im(\epsilon_\perp^{-1})$ has peak structures and sign reversals. $\Re(-\epsilon_\perp^{-1})$ has a sharp increase when frequency increases. In (b), $\Im(\epsilon_\perp^{-1})$ shows peak structure and sign reversal above $t=0$ and $\Re(-\epsilon_\perp^{-1})$ has a peak. The features broaden and move to higher temperature when frequency increases.} \label{Hall}
\end{figure}

We consider bound pair contribution to the Hall conductivity first by ignoring the terms $\chi_\text{f}^{\parallel}$ and $\chi_\text{f}^{\perp}$ in $\epsilon^{-1}$. Plots of the negative real part (blue circles) and imaginary part (red squares) of $\epsilon_\perp^{-1}$ versus (a) $\log \overline \omega$ and (b) $t$ are presented in Fig.~\ref{Hall}. Here, $\overline \omega = \omega a_0^2 /(2 \overline D)$ is the scaled frequency, and $t=T/T_\text{KT}-1$ is the reduced temperature. In this and the following figures, we use the renormalization group flow equations $\partial_l(2-X_l)^{-1} = 4\pi^2 Y_l^2$ and $\partial_l Y_l = X_l Y_l$,\cite{Ambegaokar:3} with initial conditions $X_0 = 2-[2-X_0^{(\text{KT})}]/(1+t)$ and $2\pi Y_0=0.1$ related to the bare interaction strength $K_0$ and chemical potential $\mu_0$ respectively. At transition temperature, $X_0^{(\text{KT})} = X_0(T=T_\text{KT}) \approx -0.215$ which is obtained numerically. Here, it suffices to notice that $X_l$ is given by $X_l = 2- \pi K_l$ where $ K_l = \widetilde \epsilon^{-1} K_0$ is the renormalized interaction strength, and $Y_l$ is related to the the number of bound pairs found with pair size $r=a_0 e^l$. Also, we use the convective ratio $x_0=0.1$ in the plots for illustration purpose.

We first focus on the discussion of the frequency dependence of $\epsilon_\perp^{-1}$ at some fixed temperatures in Fig.~\ref{Hall}(a). In the low temperature phase $t<0$, both $\Re(-\epsilon_\perp^{-1})$ and $\Im(\epsilon_\perp^{-1})$ are positive and increase steadily with frequency. When temperature increases to the high temperature phase $t=0.045$, $\Im(\epsilon_\perp^{-1})$ shows a positive-valued peak followed by sign reversal at higher frequency indicated by the red horizontal double-headed bar. Meanwhile, $\Re(-\epsilon_\perp^{-1})$ increases sharply around the peak of $\Im(\epsilon_\perp^{-1})$. These features move to higher frequency side when temperature increases further ($t=0.090$). For temperature dependence at fixed frequencies in Fig.~\ref{Hall}(b), the sign reversal and peak structure in $\Im(\epsilon_\perp^{-1})$ are also observed when temperature varies above $T_\text{KT}$. Meanwhile, $\Re(-\epsilon_\perp^{-1})$ shows a simple peak structure across $T_\text{KT}$. When frequency increases, the structure broadens and moves to higher temperature region.

Sign anomaly in Hall conductivity has been observed in various superconducting systems such as high-temperature SCs, \cite{Artemenko:1,Hagen:1,Hagen:2} and it is known that single vortex ``upstream" motion or change of sign of charge carriers could give rise to such phenomenon.\cite{Kopnin:3} Here in our model the sign change in $\Im(\epsilon_\perp^{-1})$ stems from the vortex-antivortex pair unbinding process. For a vortex-antivortex pair moving ``downstream" with same speed, total transverse electric field is canceled and no Hall effect can be observed. Now that the constituting vortices are not equivalent to each other, they can respond differently under the influence of transport current, and a net transverse electric field follows.

\begin{figure}[htbp]
\includegraphics [width=0.8\textwidth] {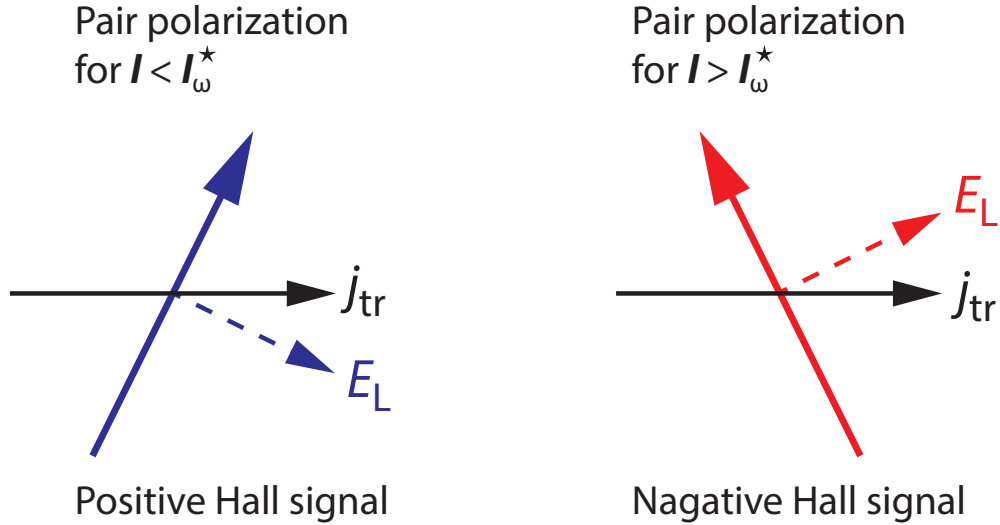}
\caption{(Color online) Schematic diagram of vortex-induced electric field $\bm E_\text{L}$ under transport current $\bm j_\text{tr}$ with convective ratio $x_0>0$. The black solid arrow is the transport current $\bm j_\text{tr}$. The blue and red solid arrow represent the pair polarization of pair with size $l<l_\omega^\star$ and $l>l_\omega^\star$ respectively. The dashed arrows show the electric field generated by pair motion in the respective cases. Pairs with size $l<l_\omega^\star$ contribute to positive $\Im(\epsilon_\perp^{-1})$ and vice verse. The net $\bm E_\text{L}$ determines the overall sign of $\Im(\epsilon_\perp^{-1})$. } \label{polarization}
\end{figure}

From the dip-recouping shape of $\Im [\mathcal G_\perp (r,\omega)]$ in Fig.~\ref{mathcalG}, the dip (recouping) region with pair size $l<l_\omega^\star$ ($l>l_\omega^\star$) contributes to positive (negative) $\Im(\epsilon_\perp^{-1})$, since $\Im(\epsilon_\perp^{-1}) \propto \Im (- \chi_\text{b}^\perp) \propto \int_{a_0}^{\xi_+} dr Y^2(r) \Im[ -\mathcal G_\perp (r,\omega)]/r$. The function $Y^2 \delta r \propto r( d\widetilde \epsilon/dr) \delta r$ is proportional to the number of pairs present in a ring with radius $r$ and width $\delta r$. Therefore, the sign of $\Im(\epsilon_\perp^{-1})$ depends on whether there are more pairs with pair size smaller or larger than $l_\omega^\star$. An interesting point to mention is that, upstream vortex motion is not required for negative $\Im(\epsilon_\perp^{-1})$; instead, it is the asymmetric pair motion in opposite angular directions in Fig.~\ref{capitalG} that brings about the Hall anomaly. We interpret such result to be that the direction of electric field $\bm E_\text{L}$ generated by a vortex pair is also pair-size-dependent. This interpretation is illustrated in Fig.~\ref{polarization}. The pair with $l<l_\omega^\star$ polarizes in a such a way that $\bm E_\text{L}$ points in the direction giving positive Hall signal and vice verse. The final sign of $\Im(\epsilon_\perp^{-1})$ depends on the net $\bm E_\text{L}$ caused by all the pairs.

\begin{figure}[htbp]
\includegraphics [width=0.8\textwidth] {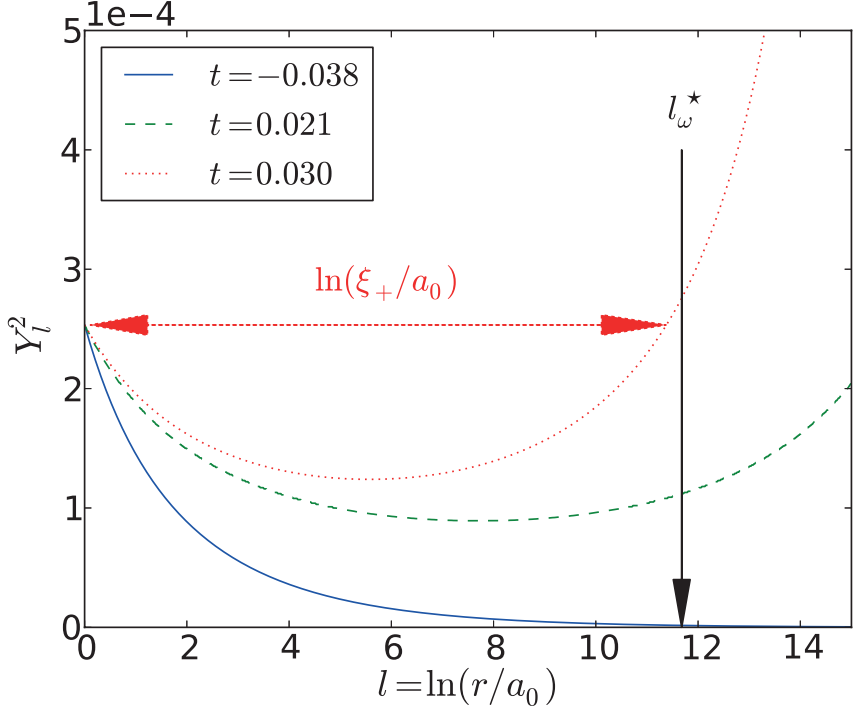}
\caption{(Color online) A plot of $Y_l^2$ versus $l$ for $t = -0.038$, $0.021$, and $0.030$. The vertical arrow marks $l_\omega^\star$, the position where $\Im (\mathcal G_\perp)$ changes sign. $2\pi Y_0 =0.1$. At $t = 0.030$, $\ln (\xi_+/a_0)$ is indicated by the red double-headed arrow. Three representative situations concerning the sign issue of $\Im(\epsilon_\perp^{-1})$ are $Y_l^2$ increasing at $l_\omega^\star$, $Y_l^2$ decreasing at $l_\omega^\star$, and $l_\omega^\star \approx \ln (\xi_+/a_0)$.} \label{Y2}
\end{figure}

The weight function $Y_l^2$, which describes the probability of a pair having pair size $r=a_0 e^l$, is plotted in Fig.~\ref{Y2} at different temperatures. The other parameters used are $\log \overline \omega = -9$, $x_0=0.1$, and $2\pi Y_0=0.1$. For $t = -0.038$, $Y_l^2$ is slightly decreasing at $l_\omega^\star$, so the dip-recouping shape in $\Im [\mathcal G_\perp (r,\omega)]$ almost cancels each other in the integral in Eq.~(\ref{chibb}), giving a weak positive $\Im(\epsilon_\perp^{-1})$. If the temperature increases, for example $t=0.021$, $Y_l^2$ becomes increasing at $l_\omega^\star$. More pairs are present in the recouping region and the sign change of $\Im(\epsilon_\perp^{-1})$ thus occurs. If the temperature increases further to $t=0.030$, the coherence length $\xi_+$ characterizing the maximum pair size becomes comparable to $l_\omega^\star$. As a result, only the dip part is integrated because the recouping part lies outside of the integration limit in Eq.~(\ref{chibb}). A strong positive-valued peak then replaces the sign reversal. In summary, temperature controls the pair size distribution, which then determines the sign of $\Im(\epsilon_\perp^{-1})$. Since the positive slope of $Y_l^2$ at $l_\omega^\ast$ only occurs at the high temperature phase, the claim that the sign reversal of $\Im(\epsilon_\perp^{-1})$ is related to KT transition is justified.

To close this subsection, we discuss the Hall conductivity in the static limit. For low temperature phase $t<0$, Hall conductivity diverges in zero-frequency limit. Log-log plots of $\Im(\epsilon_\perp^{-1})$ and $\Re(-\epsilon_\perp^{-1})$ versus $ \omega$ using linear fitting show that $\Im(\epsilon_\perp^{-1}) \propto { \omega}^{\alpha_\text{I}}$ and $\Re(-\epsilon_\perp^{-1}) \propto { \omega}^{\alpha_\text{R}}$ with both $\alpha_\text{R}$ and $\alpha_\text{I}$ greater than zero but smaller than unity for the temperature range $t =-0.037$ down to $-0.15$ and frequency range $\log \overline \omega = -10$ to $-2$. This mean that $\epsilon_{\perp}^{-1}$ decreases slower than $\omega$ when frequency approaches zero, and as a result $\sigma_{\perp} \propto \epsilon_{\perp}^{-1}/(i\omega)$ diverges at $\omega \rightarrow 0$.

\subsection{Power dissipation from bound pair contribution}

\begin{figure}[htbp]
\includegraphics [width=0.8\textwidth] {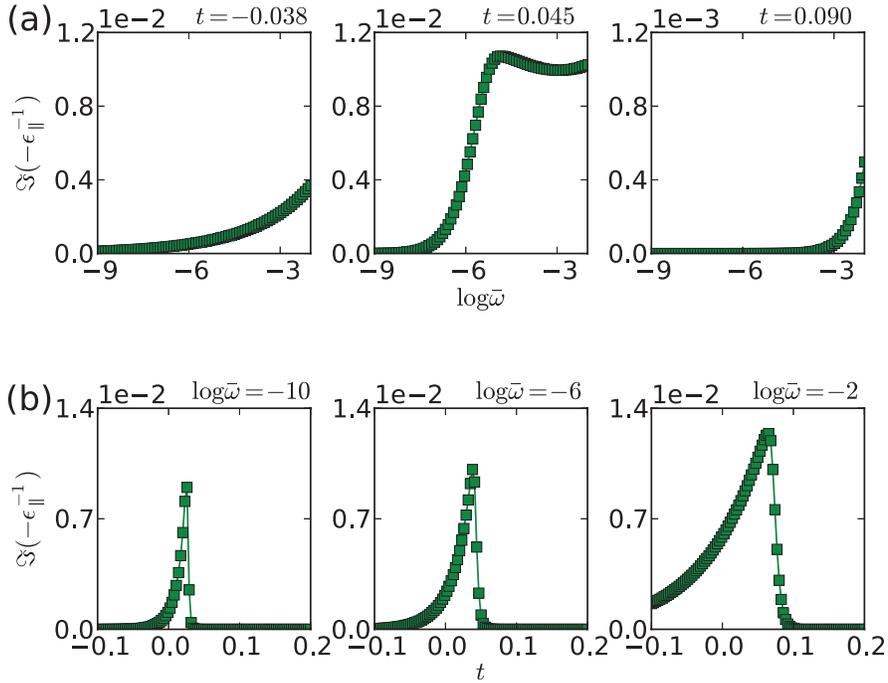}
\caption{(Color online) Bound pair contribution. $\Im(-\epsilon_\parallel^{-1})$ as a function of (a) $\log \overline \omega$ and (b) $t=T/T_\text{KT}-1$. $x_0=0.1$ and $2\pi Y_0=0.1$. In (a), $\Im(-\epsilon_\parallel^{-1})$ increases with frequency when $t<0$ and is suppressed at small frequency when $t>0$. In (b), $\Im(-\epsilon_\parallel^{-1})$ has peak structure around $T_\text{KT}$. The peak broadens and moves to higher temperature when frequency increases.}  \label{Diss}
\end{figure}

In this subsection, we discuss the frequency and temperature dependence of $\Im (-\epsilon^{-1}_\parallel)$. Again, we consider bound pair contribution here. In Fig.~\ref{Diss}, plots of $\Im(-\epsilon_\parallel^{-1})$ versus (a) $\log \overline \omega$ and (b) $t$ are presented. The other parameters used are the same as those in Fig.~\ref{Hall}. In Fig.~\ref{Diss}(a), $\Im(-\epsilon_\parallel^{-1})$ increases steadily with frequency at low temperature phase $t<0$. When $t>0$, it is suppressed at small frequency. In Fig.~\ref{Diss}(b), $\Im(-\epsilon_\parallel^{-1})$ first increases with temperature, and is suppressed above $T_\text{KT}$, resulting in a peak structure. At higher frequency, the peak is broadened and shifts to higher temperature. These features can be understood under the standard AHNS theory: Above $T_\text{KT}$, the coherence length $\xi_+$ becomes finite, and the integration limits cannot cover the peak structure in $\Im \mathcal (\mathcal G_\parallel)$ at large $l$. Thus, $\Im(-\epsilon_\parallel^{-1})$ is suppressed at low frequency for $t>0$ in Fig.~\ref{Diss}(a), and above some temperatures in Fig.~\ref{Diss}(b). Indeed, it is not surprising that the result behaves similarly to its $s$-wave counterpart when we notice that the shape of $\mathcal G_\parallel$ is qualitatively similar to that of $g_s$ even for finite $x_0$.

In zero-frequency limit in the low temperature phase, $\Im(-\epsilon_{\parallel}^{-1})$ decreases slower than $\omega$ when frequency approaches zero and thus the $\Re(\sigma_\parallel) \propto \Im(-\epsilon_{\parallel}^{-1})/\omega$ diverges. This is found in log-log plots of $\Im(-\epsilon_\parallel^{-1})$ versus $\omega$ that $\Im(-\epsilon_\parallel^{-1}) \propto { \omega}^{\beta}$ with $\beta>0$ but smaller than unity using linear fitting from $t =-0.037$ to $-0.15$ and $\log \overline \omega =-10$ to $-2$. However, we cannot interpret it to be the divergence of power dissipation because the dissipation also depends on the magnitude of the electric field in the superconducting bulk which is supposed to be vanishingly small in the static limit.

\subsection{Total conductivity tensor and free vortex contribution}

\begin{figure}[htbp]
\includegraphics [width=0.8\textwidth] {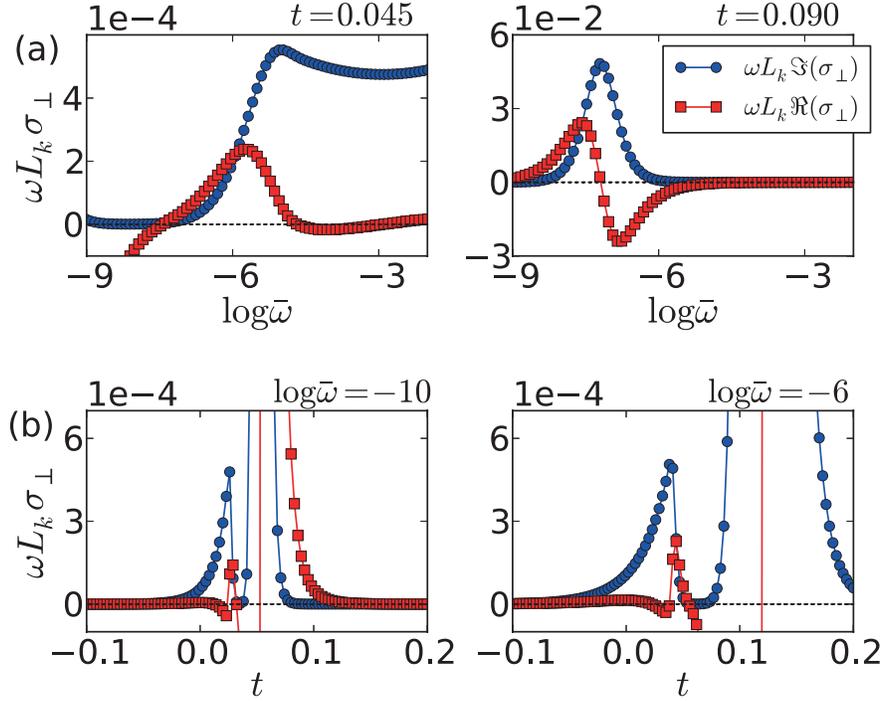}
\caption{(Color online) Total $\omega L_k \sigma_\perp$ as a function of (a) $\log \overline \omega$ and (b) $t$. The red squares and the blue circles represent the real part and imaginary part of $\omega L_k \sigma_\perp$ respectively. $C_1=1$ and $b=2$ for illustration. The other parameters used are the same as those in Fig.~\ref{Hall}.} \label{totalHall}
\end{figure}

\begin{figure}[htbp]
\includegraphics [width=0.8\textwidth] {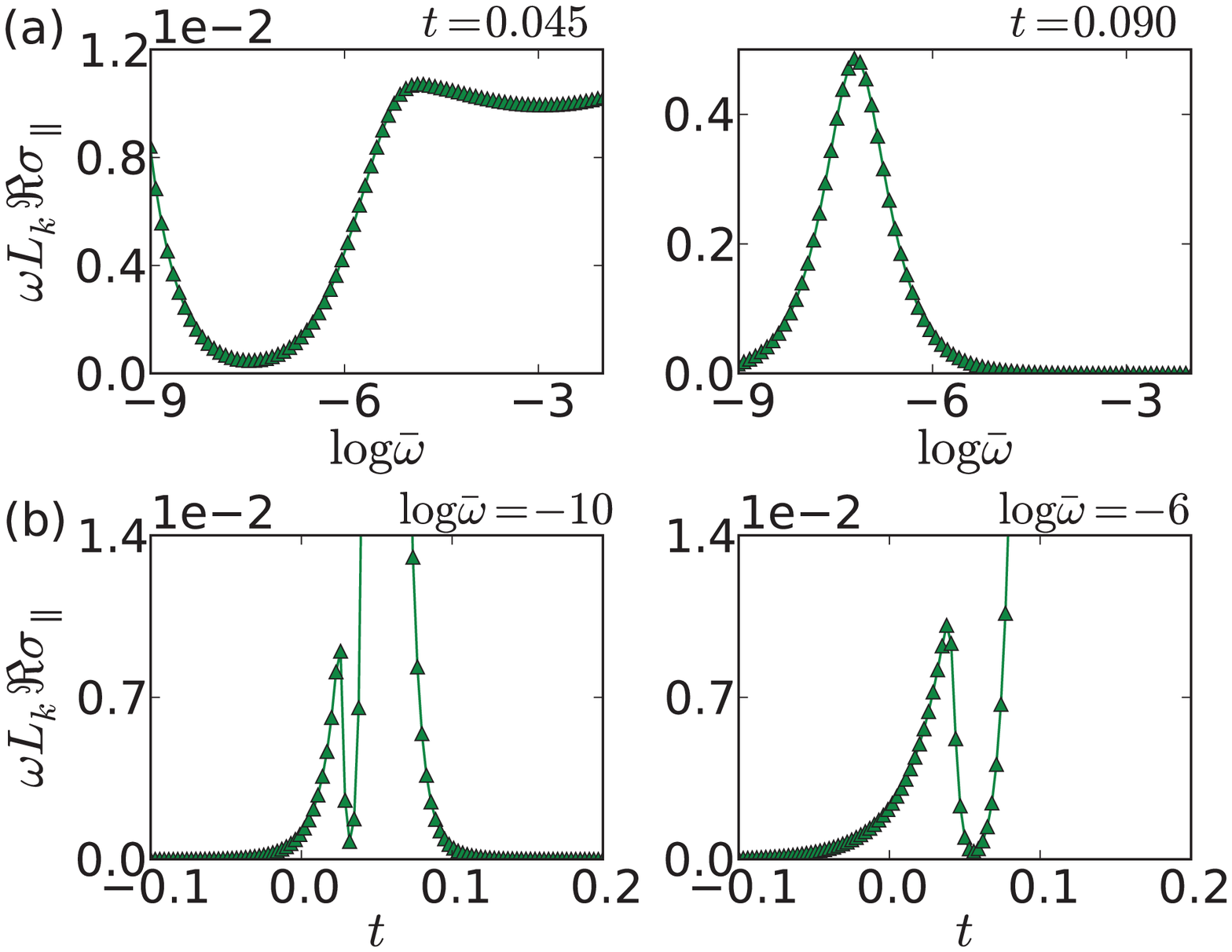}
\caption{(Color online) Total $\omega L_k \Re(\sigma_\parallel)$ as a function of (a) $\log \overline \omega$ and (b) $t$. The other parameters used are the same as those in Fig.~\ref{totalHall}.}  \label{totalDiss}
\end{figure}

\begin{figure}[htbp]
\includegraphics [width=0.8\textwidth] {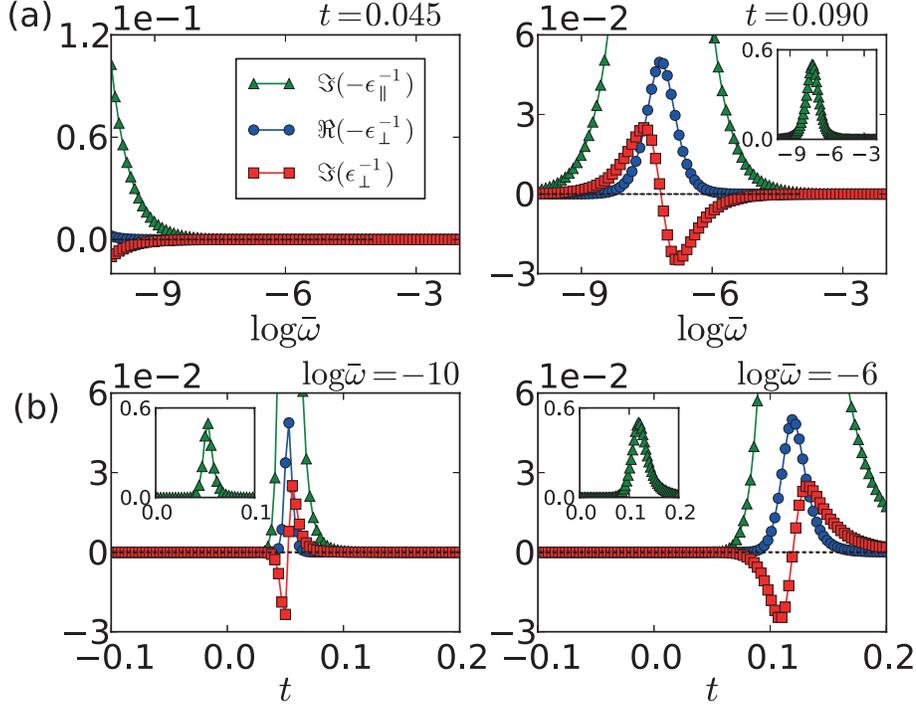}
\caption{(Color online) Free vortex contribution to $\Im (-\epsilon_\parallel^{-1})$, $\Re (-\epsilon_\perp^{-1})$, and $\Im (\epsilon_\perp^{-1})$ as a function of (a) $\log \overline \omega$ and (b) $t$. The green triangles represent $\Im (-\epsilon_\parallel^{-1})$. The blue circles and the red squares represent the negative real part and imaginary part of $\epsilon_\perp^{-1}$ respectively. The insets show the full plot range of $\Im (-\epsilon_\parallel^{-1})$. The other parameters used are the same as those in Fig.~\ref{totalHall}.}  \label{figfree}
\end{figure}

Having discussed the result due to bound pair dynamics, we study the total contribution from both bound pair and free vortex dynamics in this subsection. In order to discern the contribution of bound pair and free vortex dynamics to the total Hall conductivity and power dissipation, we plot the total conductivity tensor $\omega L_k \sigma_\perp$ and $\omega L_k \Re (\sigma_\parallel)$ in Figs.~\ref{totalHall} and \ref{totalDiss} as a function of (a) $\log \overline \omega$ and (b) $t$, as well as the $\Im (-\epsilon_\parallel^{-1})$, $\Re (-\epsilon_\perp^{-1})$, and $\Im (\epsilon_\perp^{-1})$ due purely to the free vortex motion in Fig.~\ref{figfree}. We can compare these figures with those which take only the bound pair dynamics into account (Figs.~\ref{Hall} and \ref{Diss}). For the sake of illustration, we use $C_1=1$ and $b=2$ for the weight of free vortex contribution.

For the frequency dependence in Figs.~\ref{totalHall} and \ref{totalDiss} at temperature $t=0.045$, we can see free vortex signal emerging at small frequency region. If the temperature increases further at $t=0.090$, the free vortex signal can merge with and outweigh the bound pair signal [compared with Figs.~\ref{Hall} and \ref{figfree}(a) $t=0.090$]. For $t<0$ (not shown), since there is no free vortex contribution, we expect the curves are the same as those in low temperature phase in Figs.~\ref{Hall} and \ref{Diss}(a). As for the temperature dependence in Figs.~\ref{totalHall} and \ref{totalDiss}(b), we can see for $\log \overline \omega =-10$ the free vortex signal appears at temperature very close to the bound pair signal and they almost merge into each other. When the frequency increases ($\log \overline \omega = -6$), this extra strong signal moves to higher temperature and becomes distinguishable from the bound pair signal. The magnitude of the free vortex signal can be seen in Fig.~\ref{figfree}.

As discussed in the previous two subsections, $\Re (\sigma_\perp)$, $\Im (\sigma_\perp)$, and $\Re (\sigma_\parallel)$ diverge when $\omega \rightarrow 0$. However, it is only true in low temperature phase. In the case where $t>0$, free vortex contribution dominates in the static limit $\gamma_0/ \omega \gg 1$. With such condition, we have $\epsilon \approx \chi_\text{f}$ which is purely imaginary. From Eq.~(\ref{free}), the diagonal part of $-\Im (\chi_\text{f}^{-1})$ and the off-diagonal part of $\Im (\chi_\text{f}^{-1})$ are given by $ \omega/[\gamma_0(1+x_0^2)]$ and $x_0 \omega/[\gamma_0(1+x_0^2)]  $ respectively, and therefore both of them are proportional to $\omega \xi_+^2$. This shows that $\sigma \propto \chi_\text{f}^{-1}/(-i\omega)$ is purely real, constant in frequency, and diverges when $T \rightarrow T_\text{KT}^+$. As a result, in the high temperature phase, both the Hall conductivity and dissipation are protected from divergence in the zero-frequency limit but have strong signal when approaching $T_\text{KT}$ from above.

Finally, Fig.~\ref{figfree} shows $\Im (-\epsilon_\parallel^{-1})$, $\Re (-\epsilon_\perp^{-1})$, and $\Im (\epsilon_\perp^{-1})$ as a function of (a) $\log \overline \omega$ and (b) $t$ including only the free vortex but not the bound pair contribution. The frequency and temperature range are chosen to match those in Figs.~\ref{totalHall} and \ref{totalDiss}. By Eq.~(\ref{free}), we see that important features in the figure appear around the crossover from $\gamma_0/\omega \gg 1$ to the opposite limit $\gamma_0/\omega \ll 1$. In Fig.~\ref{figfree}(a) on the right panel, when observing from small frequency region $\gamma_0/\omega \gg 1$, we see that all three quantities increase with frequency. When frequency increases further, $\Im (\epsilon_\perp^{-1})$ changes sign and then all the quantities decrease in magnitude with frequency, which means that it starts to be inefficient for the free vortex to respond to the perturbation when frequency is high. In Fig.~\ref{figfree}(b), similar features can be observed in the temperature domain. Besides, we can see that the main features in the curves emerge at a temperature closer to $T_\text{KT}$ for small frequency (left panel) than for high frequency (right panel). Although the sign change in $\Im (\epsilon_\perp^{-1})$ can also be observed in free vortex picture, it does not arise from the response function $\mathcal G_\perp$ mentioned before. Instead, it is attributed to the pole structure $1+ i\gamma_0/\omega =0$.

\section{summary and remark}

In this work, we have generalized AHNS's vortex dynamics in the chiral $p$-wave superconducting state. The behavior of the conductivity tensor near KT transition is investigated. We show that the Hall conductivity can be nonzero arising from the vortex-antivortex pair unbinding process or from free vortex motion even in the absence of magnetic field. Power dissipation is also predicted in this dynamical picture. In low temperature phase, both the $\sigma_\perp$ and $\Re(\sigma_\parallel)$ diverge in the static limit. But in high temperature phase, the contribution from free vortex motion gives finite results in the static limit. We can distinguish the bound pair and free vortex contribution to the total conductivity tensor by comparing the figures from different contribution.

Sign reversal and strong positive peak in $\omega \Re(\sigma_\perp)$ are illustrated in some suitable frequency close to transition, as well as above the transition temperature at fixed frequencies. On the other hand, the $\omega \Re(\sigma_\parallel)$ behaves in a fashion similar to that of the $s$-wave case. In bound pair description, the induced Hall conductivity is strongly influenced by the dip-recouping shape of transverse response function $\Im (\mathcal G_\perp)$, which stems from the asymmetric angular response of pairs to the driving field. The convective term in the Fokker-Planck equation, which originates from the broken $\mathcal T$ symmetry nature of a chiral $p$-wave SC, is essential to this asymmetry. All these results depend solely on the convective ratio $x_0$, and are valid even without applied magnetic field.

In our work, temperature dependence of the drag coefficients is omitted. Theoretical calculation of the drag coefficients has been performed microscopically for SCs \cite{Kopnin:1} and for SF $^3$He (Refs.~\onlinecite{Kopnin:2,Kopnin:4}) at temperatures considerably lower than the superconducting transition temperature by the use of quasiclassical theory. Generalization of such calculation in chiral $p$-wave pairing state around $T_\text{KT}$ will be useful to estimate the strength of the convective ratio. Indeed, if we use the formulas of drag coefficients which only take into account the Caroli-deGennes-Matricon mode at very low temperature
$
\eta^{(i)} = (2\pi \hbar \rho_s^0 /m^\ast)\omega_0 \tau^{(i)}/[(\omega_0 \tau^{(i)})^2 + 1]
$ and $
{\eta^\prime}^{(i)} = (2\pi \hbar \rho_s^0 /m^\ast)(\omega_0 \tau^{(i)})^2/[(\omega_0 \tau^{(i)})^2 + 1 ]
$, \cite{Kopnin:1}
we can arrive at the result $D^{(i)} = k_\text{B}T m^\ast /(2\pi \hbar \rho_s^0 \omega_0 \tau^{(i)})$ and $C^{(i)}=0$, which results in $x_0=0$. However, we believe that the effect of temperature and delocalized excitations can contribute to the convective ratio.

\appendix

\section{Appendix}

In this Appendix, we apply certain approximation to obtain analytic expression of $\Im(-\epsilon_\parallel^{-1})$, $\Im(\epsilon_\perp^{-1})$, and $\Re(-\epsilon_\perp^{-1})$ in the bound pair dynamics picture.

\subsection{Analytic expression of susceptibility matrix elements}

We first obtain analytic expression for susceptibility matrix elements $\chi_\text{b}^{\parallel}$ and $\chi_\text{b}^{\perp}$ by employing an approximate solution
\begin{eqnarray}
g_\pm (r,\omega) \approx \frac{1}{ 1 -i\omega r^2/(\lambda \overline D) \pm i x_0 }
\end{eqnarray}
and taking $r (d \widetilde \epsilon/dr) $ to be a constant function of $r$. We rewrite Eq.~(\ref{chibb}) in the form
\begin{eqnarray}
\chi_\text{b}^{\parallel}
&=&
\int_{a_0}^{\xi_+} dr \frac{d \widetilde \epsilon}{dr}
\left[
\left( \frac{ g_+ + g_-}{2} \right)
+
 ix_0
\left( \frac{ g_+ - g_-}{2} \right)
\right], \label{chi_para0}
\\
\chi_\text{b}^{\perp}
&=&
-i\int_{a_0}^{\xi_+} dr \frac{d \widetilde \epsilon}{dr}
\left[
\left( \frac{ g_+ - g_-}{2} \right)
+
ix_0
\left( \frac{ g_+ + g_-}{2} \right)
\right], \label{chi_perp0}
\end{eqnarray}
and further approximate the two functions appearing in the above integrands by
\begin{eqnarray}
\frac{ g_+ + g_-}{2}
&\approx &
\left[
\left(
\frac{  1}{ 1+x_0^2}\right) \theta (r_4 - r)
+
\text{sgn}(x_0) r I_1 \delta(r-r_1)
\right]
\nonumber\\
&&
+
i \left[r I_2 \delta(r-r_2) - x_0 r I_1 \delta(r-r_1)\right] ,
\\
\frac{ g_+ - g_-}{2}
&\approx &
r I_1 \delta(r-r_1)
+ i r I_3 \delta(r-r_3)
\nonumber\\
&&
-ix_0
\left[
\left(
\frac{  1}{ 1+x_0^2}\right) \theta (r_4 - r)
+
\text{sgn}(x_0) r I_1 \delta(r-r_1)
\right],
\end{eqnarray}
where
\begin{eqnarray}
I_1 &\equiv&
\frac{1}{2} \int_0^\infty dr \frac{1}{r}
\left(
\frac{1}
{1+ [ \omega r^2 / (\lambda \overline D)- x_0]^2 }
-
\frac{1}
{1+ [ \omega r^2 / (\lambda \overline D)+ x_0]^2 }
\right)
\nonumber\\
&=&
 \frac{ x_0  \pi}
{4+4 x_0^2},
\\
I_2 &\equiv& \frac{1}{2} \int_0^\infty dr \frac{1}{r}
\left(
\frac{\omega r^2 / (\lambda \overline D)}
{1+ [\omega r^2 / (\lambda \overline D)- x_0]^2 }
+
\frac{\omega r^2 / (\lambda \overline D)}
{1+ [\omega r^2 / (\lambda \overline D)+ x_0]^2 }
\right)
\nonumber\\
&=&
\frac{ \pi} {4},
\\
I_3 &\equiv&
\frac{1}{2} \int_0^\infty dr \frac{1}{r}
\left(
\frac{\omega r^2 / (\lambda \overline D)}
{1+ [\omega r^2 / (\lambda \overline D)- x_0]^2 }
-
\frac{\omega r^2 / (\lambda \overline D)}
{1+ [\omega r^2 / (\lambda \overline D)+ x_0]^2 }
\right)
\nonumber\\
&=&
\frac{1}{4}
\left[
\cot^{-1} \frac{2 x_0}{-1+x_0^2} + \frac{\pi}{2} \text{sgn}(x_0)
\right].
\end{eqnarray}

In the above expression, $r_1$, $r_2$, $r_3$, and $r_4$ are some characteristic lengths defined as follows: $r_1$, $r_2$, and $r_3$ are the peak position of the integrand of $I_1$, $I_2$, and $I_3$ respectively. $r_4$ is chosen such that $1/ \{1+ [ \omega r_4^2 / (\lambda \overline D)+ x_0]^2 \} = 1/ [2(1+ x_0^2 )]$. They are given respectively by
$
r_1^2 =
 \lambda \overline D (
-1 + x_0^2
+  2 \sqrt{1+x_0^2+x_0^4}
)^{1/2}/ (\sqrt{3} \omega)
$, $
r_2^2=r_3^2 =
 \lambda \overline D \sqrt{1+x_0^2} /  \omega
$, and
$
r_4^2 = \lambda \overline D (\sqrt{1+2x_0^2} - x_0)/ \omega
$. One more characteristic length worth mentioning is the pair size $r_\omega^\star$ at which $\Im (\mathcal G_\perp)$ changes its sign. Within the approximation described above, we find that $r_\omega^\star= r_2 = r_3$. Then we can obtain the quantity $l_\omega^\star = \ln (r_\omega^\star/a_0)$ mentioned in the main text.

Finally, using $r d \widetilde \epsilon /dr = 4\pi^3 K_0 Y^2 (r)$, we obtain analytic expression for the components of the susceptibility matrix for $\xi_+ > r_j$
\begin{eqnarray}
\chi_\text{b}^{\parallel}
&\approx&
\widetilde \epsilon_4 -1
+
\text{sgn}(x_0) x_0 \pi^4 K_0 Y_1^2
\nonumber\\
&&
-
x_0
\left(
\cot^{-1} \frac{2 x_0}{-1+x_0^2} + \frac{\pi}{2} \text{sign }x_0
\right) \pi^3 K_0 Y_3^2
+
i \pi^4 K_0 Y_2^2, \label{chi_para}
\\
\chi_\text{b}^{\perp}
&\approx&
\left(
\cot^{-1} \frac{2 x_0}{-1+x_0^2} + \frac{\pi}{2} \text{sign }x_0
\right)
\pi^3 K_0 Y_3^2
+
i x_0  \pi^4 K_0 (Y_2^2 -Y_1^2), \label{chi_perp}
\end{eqnarray}
where some shorthand notations are introduced: $\widetilde \epsilon_4 = \widetilde \epsilon(r_4)$ and $Y_j = Y(r_j)$. From the expression above, we notice that $\chi^{\parallel}_\text{b}$ ($\chi^{\perp}_\text{b}$) is (anti-) symmetric in $x_0$ as it should be. We can also identify the contribution from the dip and recouping part of $\Im (\mathcal G_\perp)$ to be $Y_1^2$ and $Y_2^2$ in the imaginary part of Eq.~(\ref{chi_perp}).

\subsection{Analytic expression of $\Im(-\epsilon_\parallel^{-1})$, $\Im(\epsilon_\perp^{-1})$, and $\Re(-\epsilon_\perp^{-1})$}

In this subsection, $\Im(-\epsilon_\parallel^{-1})$, $\Im(\epsilon_\perp^{-1})$, and $\Re(-\epsilon_\perp^{-1})$ for nonzero $x_0$ are calculated in two limiting cases $x_0 \ll1$ and $x_0 \gg1$. For $x_0 \ll1$, we take $r_4^2 \approx r_2^2=r_3^2 \approx \lambda \overline D/ \omega$, $r_1^2 \approx \lambda \overline D/(\sqrt{3} \omega )$, and
\begin{eqnarray}
\cot^{-1} \left( \frac{2 x_0}{-1+x_0^2} \right) + \frac{\pi}{2} \text{sign }x_0
 \approx 2 x_0 - \frac{2x_0^3}{3} + \frac{2 x_0^5}{5} - \dotsm.
\end{eqnarray}
At the opposite limit $x_0 \gg 1$, we take $r_4^2 \approx \lambda \overline D x_0 (-1+\sqrt{2}) / \omega$, $r_1^2 \approx r_2^2=r_3^2 \approx \lambda \overline D x_0/ \omega$, and
\begin{eqnarray}
\cot^{-1} \left(\frac{2 x_0}{-1+x_0^2} \right) + \frac{\pi}{2} \text{sgn}(x_0)
&\approx& \pi \text{sgn}(x_0) -\frac{2}{x_0} +\frac{2}{3x_0^3}  -\frac{2}{5x_0^5} + \dotsm.
\end{eqnarray}
Besides, we neglect the $x_0$ dependence of $\widetilde \epsilon_4$ and $Y_j$ in the expansion. Such approximation is valid for $x_0 \ll 1$ and all temperature near $T_\text{KT}$ because $r_j$ is almost independent of $x_0$. It is also valid for $x_0 \gg 1$ in low temperature phase because the renormalization almost go to completion at large $x_0$ and $Y_j$ can be treated as independent of $x_0 \propto r_j^2$. But it is not valid for $x_0 \gg 1$ in high temperature phase for two reasons: first, $Y_j$ is increasing at large $x_0$ and second, $r_j$ are very likely to be greater than $\xi_+$ so that the integrations in Eqs.~(\ref{chi_para0}) and (\ref{chi_perp0}) give the trivial results $\chi_\text{b}^\parallel = \widetilde \epsilon (\xi_+) -1$ and $\chi_\text{b}^\perp = 0$.

Expanding using small $x_0$, the imaginary part of $-\epsilon^{-1}_\parallel $ is approximately given by
\begin{eqnarray}
\Im (- \epsilon^{-1}_\parallel )
&\approx&
\frac{K_0 \pi^4 Y_4^2}
{\widetilde \epsilon_4^{2} + K_0^2 \pi^8 Y_4^4}
-\vert x_0 \vert
\frac{2 \widetilde \epsilon_4 K_0^2 \pi^8 Y_4^2 Y_1^2}
{(\widetilde \epsilon_4^{2} + K_0^2 \pi^8 Y_4^4)^2}
\nonumber\\
&\approx&
\widetilde \epsilon_4^{-2} K_0 \pi^4 Y_4^2
(1-2 \vert x_0 \vert \widetilde \epsilon_4^{-1} K_0 \pi^4 Y_1^2),\label{power_app}
\end{eqnarray}
where we have used $\widetilde \epsilon_4 \gg K_0 \pi^4 Y_4^2$ in the last step. The presence of the second term acts as a suppression of dissipation when $x_0$ is switched on.  If we neglect the correction term proportional to $\vert x_0 \vert$, the expression reduces to the $s$-wave SF/SCs result. In the opposite limit $x_0 \gg1$, the imaginary part of $-\epsilon^{-1}_\parallel $ to the leading order in the smallness of $1/x_0$ is given by
\begin{eqnarray}
\Im (- \epsilon^{-1}_\parallel )
 &\approx&
\frac{ K_0 \pi^4 Y_1^2}
{( \widetilde \epsilon_4  + 2K_0 \pi^3 Y_1^2)^2 + 4K_0^2 \pi^8 Y_1^4 }
\nonumber\\
&\approx&
\widetilde \epsilon_4^{-2} K_0 \pi^4 Y_1^2,
\end{eqnarray}
assuming $\widetilde \epsilon_4  \gg 2K_0 \pi^3 Y_1^2$. Surprisingly, the leading order of $\Im(-\epsilon_\parallel^{-1})$ at large $x_0$ has the same form as its small $x_0$ counterpart. The only difference is that the characteristic length is changed from $\sqrt{\lambda \overline D / \omega}$ to $\sqrt{\lambda \overline C / (2\pi K_0\omega)}$. This can be understood if we recall that the shape of $\Im (\mathcal G_\parallel)$ has no qualitative change for finite $x_0$.

Following similar procedure we can obtain analytic expression for $\Im (\epsilon^{-1}_\perp)$ and $\Re (-\epsilon^{-1}_\perp)$. For small $x_0$, we obtain $\Im (\epsilon^{-1}_\perp)$ to the first order in $x_0$
\begin{eqnarray}
&&
\Im (\epsilon^{-1}_\perp)
\nonumber\\
&\approx&
\frac{
x_0 \left[\widetilde \epsilon_4^2 K_0 \pi^4
\left(Y_1^2 - Y_4^2\right)
+
4 \widetilde \epsilon_4 K_0^2 \pi^7 Y_4^4
+
K_0^3 \pi^{12} Y_4^6
-
K_0^3 \pi^{12} Y_4^4 Y_1^2\right]
   }
{(\widetilde \epsilon_4^2 +
  K_0^2 \pi^8 Y_4^4)^2}
\nonumber\\
&\approx&
x_0 \widetilde \epsilon_4^{-2} K_0 \pi^4
\left[
\left(Y_1^2 - Y_4^2\right)
+
4 \widetilde \epsilon_4^{-1} K_0 \pi^3 Y_4^4
\right].
\end{eqnarray}
We have used the condition $\widetilde \epsilon_4 \gg K_0 \pi^4 Y_j^2$ in the last line. From the expression, the behavior depends on whether $Y_1^2 - Y_4^2$ or $4 \widetilde \epsilon_4^{-1} K_0 \pi^3 Y_4^4$ dominates. From numerical results, $Y_1^2 - Y_4^2$ dominates at a wide range of temperature in the low temperature phase. Only when approaching the transition temperature from below and away from the vicinity $\omega \xi_\pm^2 / (\lambda \overline D) \approx 1$, the $Y_4^4$ becomes more significant. On the other hand,
\begin{eqnarray}
&&\Re (-\epsilon^{-1}_\perp)
\nonumber\\
&\approx&
\frac
{
2 x_0 (\widetilde \epsilon_4^2 K_0 \pi^3 Y_4^2 + \widetilde \epsilon_4 K_0^2 \pi^8 Y_4^4 - K_0^3 \pi^{11} Y_4^6 -
   \widetilde \epsilon_4 K_0^2 \pi^8 Y_4^2 Y_1^2)
}
{
(\widetilde \epsilon_4^2 + K_0^2 \pi^8 Y_4^4)^2
}
\nonumber\\
&\approx&
2 x_0 \widetilde \epsilon_4^{-2} K_0 \pi^3 Y_4^2 .
\end{eqnarray}
We have used $\widetilde \epsilon_4 \gg K_0 \pi^5 Y_j^2$ in the second line. For large $x_0$, we assume $\widetilde \epsilon_4 \gg 4 K_0 \pi^3 Y_1^2$ and obtain to the leading order
\begin{eqnarray}
&&\Im (\epsilon^{-1}_\perp)
\nonumber\\
&\approx&
\text{sgn}(x_0)
\frac
{
2 K_0^2 \pi^8 Y_1^4.
}
{
(\widetilde \epsilon_4 + 2 K_0 \pi^3 Y_1^2) (\widetilde \epsilon_4^2 + 4 \widetilde \epsilon_4 K_0 \pi^3 Y_1^2 +
   4 K_0^2 \pi^6 Y_1^4 + 4 K_0^2 \pi^8 Y_1^4)
}
\nonumber\\
&\approx&
\text{sgn}(x_0)2 K_0^2 \pi^8 \widetilde \epsilon_4^{-3} Y_1^4,
\end{eqnarray}
and
\begin{eqnarray}
&&\Re (-\epsilon^{-1}_\perp)
\nonumber\\
&\approx&
\text{sgn} (x_0)\frac
{
K_0 \pi^4 Y_1^2
}
{
\widetilde \epsilon_4^2 + 4 \widetilde \epsilon_4 K_0 \pi^3 Y_1^2 + 4 K_0^2 \pi^6 Y_1^4 +
 4 K_0^2 \pi^8 Y_1^4
}
\nonumber\\
&\approx&
\text{sgn} (x_0)
\widetilde \epsilon_4^{-2}  K_0 \pi^4 Y_1^2.
\end{eqnarray}
It has to be mentioned that, although we have used the condition $\widetilde \epsilon_4 \gg K_0 \pi^4 Y_j^2$ extensively, it is only valid in the temperature range $t\lesssim 0.035$. Above such temperature, numerical results show that $K_0 \pi^4 Y_j^2$ can be greater than $\widetilde \epsilon_4$ at small frequency region.

%\begin{references}


\begin{thebibliography}{99}

% Hall effect and polar Kerr effect
\bibitem{Volovik:1} G. E. Volovik, JETP {\bf 67}, 1804 (1988).

\bibitem{Furusaki:1} A. Furusaki, M. Matsumoto, and M. Sigrist, Phys. Rev. B {\bf 64}, 054514 (2001).

\bibitem{Goryo:1} J. Goryo, Phys. Rev. B {\bf 78}, 060501(R) (2008).

\bibitem{Lutchyn:1} R. M. Lutchyn, P. Nagornykh, and V. M. Yakovenko, Phys. Rev. B {\bf 80}, 104508 (2009).

\bibitem{Taylor:1} E. Taylor and C. Kallin, Phys. Rev. Lett. {\bf 108}, 157001 (2012).

\bibitem{Wysokinski:1} K. I. Wysoki\'nski, J. F. Annett, and B. L. Gy\"orffy, Phys. Rev. Lett. {\bf 108}, 077004 (2012).

\bibitem{Xia:1} J. Xia, Y. Maeno, P. T. Beyersdorf, M.M. Fejer, and A. Kapitulnik, Phys. Rev. Lett. {\bf 97}, 167002 (2006).






% KT transition and vortex dynamics
\bibitem{Barber:1} M. N. Barber, Phys. Rep. {\bf 59}, 375 (1980).

\bibitem{Kosterlitz:1} J. M. Kosterlitz and D. J. Thouless, J. Phys. C {\bf 6}, 1181 (1973). 

\bibitem{Herb:1} See, e.g., J. A. Herb and J. G. Dash, Phys. Rev. Lett. {\bf 29}, 846 (1972), and references therein.

\bibitem{Ambegaokar:1} V. Ambegaokar, B. I. Halperin, D. R. Nelson, and E. D. Siggia, Phys. Rev. Lett. {\bf 40}, 783 (1978).

%\bibitem{Ambegaokar:1} V. Ambegaokar \emph{et al}., Phys. Rev. Lett. {\bf 40}, 783 (1978).

\bibitem{Ambegaokar:2} V. Ambegaokar and S. Teitel, Phys. Rev. B {\bf 19}, 1667 (1979).

\bibitem{Ambegaokar:3} V. Ambegaokar, B. I. Halperin, D. R. Nelson, and E. D. Siggia, Phys. Rev. B {\bf 21}, 1806 (1980).

%\bibitem{Ambegaokar:3} V. Ambegaokar \emph{et al}., Phys. Rev. B {\bf 21}, 1806 (1980).

\bibitem{Vinen:1} W. F. Vinen, Prog. Low Temp. Phys. {\bf 3}, 1 (1961).

\bibitem{Kosterlitz:2} J. M. Kosterlitz, J. Phys. C {\bf 7}, 1046 (1974).

\bibitem{Bishop:1} D. J. Bishop and J. D. Reppy, Phys. Rev. Lett. {\bf 40}, 1727 (1978).

\bibitem{Festin;Festin} See, e.g., \"O. Festin \emph{et al}., Phys. Rev. Lett. {\bf 83}, 5567 (1999); \"O. Festin, P. Svedlindh, F. R\"onnung, and D. Winkler, Phys. Rev. B {\bf 70}, 024511 (2004).

%\bibitem{Festin;Festin} See, e.g., \"O. Festin \emph{et al}., Phys. Rev. Lett. {\bf 83}, 5567 (1999); \"O. Festin \emph{et al}., Phys. Rev. B {\bf 70}, 024511 (2004).

%\bibitem{Maeno:1} Y. Maeno \emph{et al}., J. Phys. Soc. Jpn. {\bf 81}, 011009 (2012).

\bibitem{Maeno:1} Y. Maeno, S. Kittaka, T. Nomura, S. Yonezawa, and K. Ishida, J. Phys. Soc. Jpn. {\bf 81}, 011009 (2012).
    
\bibitem{Volovik:2} G. E. Volovik, \emph{the Universe in a Helium Droplet} (Oxford University Press, New York, 2003).

%\bibitem{Herland;Bauer} E. V. Herland \emph{et al}., Phys. Rev. B {\bf 85}, 024520 (2012); B. Bauer \emph{et al}., Phys. Rev. B {\bf 87}, 014503 (2013).

\bibitem{Herland;Bauer} E. V. Herland \emph{et al}., Phys. Rev. B {\bf 85}, 024520 (2012); B. Bauer, R. M. Lutchyn, M. B. Hastings, and M. Troyer, Phys. Rev. B {\bf 87}, 014503 (2013).




%types of integer vortex in chiral pwave SC, they are inequivalent
\bibitem{Heeb:1} R. Heeb and D. F. Agterberg, Phys. Rev. B {\bf 59}, 7076 (1999).

\bibitem{Matsumoto:1} M. Matsumoto and R. Heeb, Phys. Rev. B {\bf 65}, 014504 (2001).

\bibitem{Kato:2} Y. Kato and N. Hayashi, J. Phys. Soc. Jpn. {\bf 70}, 3368 (2001).

\bibitem{Kato:1} Y. Kato, J. Phys. Soc. Jpn. {\bf 69}, 3378 (2000).

\bibitem{Kato:3} Y. Kato and N. Hayashi, J. Phys. Soc. Jpn. {\bf 71}, 1721 (2002).

\bibitem{Kopnin:1} N. Kopnin, \emph{Theory of Nonequilibrium Superconductivity } (Oxford University Press, New York, 2001).

\bibitem{Caroli:1} C. Caroli, P. G. de Gennes, and J. Matricon, Phys. Lett. {\bf 9}, 307 (1964).


\bibitem{Kopnin:3} N. B. Kopnin, Rep. Prog. Phys. {\bf 65}, 1633 (2002).




%relation of dielectric function and conductivity tensor
\bibitem{Halperin:1} B. I. Halperin and D. R. Nelson, J. Low Temp. Phys. {\bf 36}, 599 (1979).

\bibitem{Minnhagen:1} P. Minnhagen, Rev. Mod. Phys. {\bf 59}, 1001 (1987).



%anomalous Hall effect

\bibitem{Artemenko:1} S. N. Artemenko, I. G. Gorlova, and Y. I. Latyshev, Phys. Lett. A {\bf 138}, 428 (1989).

%\bibitem{Hagen:1} S. J. Hagen \emph{et al.}, Phys. Rev. B {\bf 41}, 11630 (1990).

\bibitem{Hagen:1} S. J. Hagen, C. J. Lobb, R. L. Greene, M. G. Forrester, and J. H. Kang, Phys. Rev. B {\bf 41}, 11630 (1990).
    
\bibitem{Hagen:2} S. J. Hagen \emph{et al.}, Phys. Rev. B {\bf 47}, 1064 (1993).





%calculate drag coefficient
\bibitem{Kopnin:2} N. B. Kopnin and M. M. Salomaa, Phys. Rev. B {\bf 44}, 9667 (1991).

\bibitem{Kopnin:4} N. B. Kopnin, Phys. Rev. B {\bf 47}, 14354 (1993).





%Others
\bibitem{Donnelly:1} R. J. Donnelly, \emph{Quantum vortices in Helium II} (Cambridge University Press, Cambridge, England, 1991).

\bibitem{Adams:1} P. W. Adams and W. I. Glaberson, Phys. Rev. B {\bf 35}, 4633 (1987).


\end{thebibliography}
\end{document}